# The VOI-Based Valleytronics

G. Y. Wu,[1,2,*] N.-Y. Lue,[1] and Y.-C. Chen,[1]

Abstract

We discuss the valley-orbit interaction (VOI) and the concept of VOI based valleytronics. Potential of such valleytronics is illustrated, with graphene as an example material, in several frontier applications comprising FETs, quantum computing, and quantum communications. Two important devices are discussed as examples, namely, 1) valley pair qubits in coupled graphene quantum dots, to build quantum networks consisting of graphene and photons, and 2) valley-based FETs consisting of graphene quantum wires (channels) and armchair graphene nanoribbons (sources and drains), to build graphene-based, low-power FET circuits. This demonstration makes the VOI-based valleytronics an attractive R & D direction in the area of electronics.

[1] Department of Physics, National Tsing-Hua University, Hsin-Chu, Taiwan 30013

[2] Department of Electrical Engineering, National Tsing-Hua University, Hsin-Chu, Taiwan 30013

[*] Email: yswu@ee.nthu.edu.tw

## I. Introduction

The emerging graphene[1] has provided an exciting continent of physics for exploration[1-4] and, as a promising future electronic material, numerous interesting possibilities for frontier applications. In particular, it has opened the door to implement a new realm of electronics, namely, valleytronics, based on the valley degree of freedom[5] in electrons for the control of electrical properties[5-16]. Valley filtering (or polarization)[5-12,15], magnetic effects[6,7], and devices[5,9,13-16] have been explored or proposed.

Graphene is a hexagonal, two-dimensional crystal consisting of two identical sublattices (A and B) of carbon atoms, with conical conduction and valence bands touching at the so-called Dirac points, located at K and K' of the Brillouin zone. Because the two energy valleys of the conduction band at K and K' are degenerate and independent, electrons in graphene carry the novel, binary-valued degree of freedom (d.o.f.) – valley, in addition to spin and charge. Among prototype devices, valley qubits[13,14] or valley FETs[16] have been theoretically demonstrated. As a result, this expands the family of graphene devices which are charge- [1,17] or spin-[18] based, and adds great flexibility to the device design in electronics.

There is a strong resemblance between the valley d.o.f. and the spin d.o.f., which is more so in gapped graphene. For example, the states ({K, K'} in the valley case or {↑, ↓} in the spin case constitute a time-reversal symmetric pair of degenerate states, both of which can be split by a magnetic field. In the valley case, as shown previously, K and K' electrons in gapped graphene carry finite but opposite magnetic moments.[6,7] These moments interact with the magnetic field normal to the graphene plane, with the interaction given by

$$H_Z = \tau \mu_v B_{normal},$$

($\tau = +/-$ being the valley index for K / K', and $\mu_v$ = valley magnetic moment), which is analogous to the Zeeman interaction between a spin and a magnetic field, and breaks the valley symmetry resulting in valley splitting. Apart from $H_Z$, there is also the coupling between valley and orbit, called the valley-orbit interaction (VOI) and given by[13,14,16,19,]

$$H_{vo} = \tau \frac{\hbar}{4m^*\Delta}(\nabla V \times \hat{p})_z$$

($2\Delta$ = energy gap, $m^*$ (effective mass) = $\Delta / v_F^2$, $v_F$ = Fermi velocity, $V$ = electric potential, $p$ = momentum operator, and the subscript "z" = the Cartesian component normal to graphene plane), which is the analogy of the spin-orbit interaction (SOI).

As far as spintronics is concerned, graphene is a material with weak SOI[4], which makes the implementation of SOI-based spintronics in graphene a challenging task. The existence of the SOI's analogue – VOI in graphene is therefore of particular interest from the application point of view. Just as the SOI is the physical mechanism with a potential to realize the electrical control of electron spins[20,21], so is the VOI for valley manipulation. In regard to the application of VOI, several important conclusions have been drawn in previous studies[13,14,16], as enumerated below. (I) It has a large strength. (II) It enables electrical control of the valley d.o.f. with standard electrical gate operations. (III) It differs from the SOI in the aspect of scattering – the VOI is valley-conserving and does not cause flip type scattering. These facts make the VOI-based valleytronics an attractive R & D direction in the area of electronics.

The concept of VOI-based valleytronics, although derived from graphene, can be realized generally in a similar material system (e.g., silicene[22]) characterized by i) a hexagonal lattice with a π-electron on each lattice point, and ii) a massive Dirac particle-like electron spectrum. Condition i) ensures the corresponding Brillouin zone to be hexagonal with the existence of two degenerate and inequivalent valleys at K and K' points of the Brillouin zone. Condition ii) allows the patterning of quantum wires or dots in the material with electrical gates, and also the presence of $H_{vo}$ in the material.

In this work we focus on the material of graphene and illustrate such valleytronics with two important examples – quantum networks[13,14] and FETs[16], with emphasis on the principles underlying the devices. For the application of FETs, we illustrate an all-graphene, valley-based FET consisting of a graphene quantum wire (QW) (channel) and two armchair graphene nanoribbons (AGNRs) (source and drain), for building low-power FET circuits in graphene. For the application of quantum networks, we demonstrate a valley-based qubit made of coupled quantum dots (QDs) in graphene, and discuss how the structure may be exploited for both quantum computing and quantum communications, to build quantum networks[23] consisting primarily of graphene and photons. In both applications, the devices envisioned are gated structures, with the potential to be fabricated into integrated circuits and to be fully controlled through electrical gates – a mode of operation which has proven a great success in Si integrated circuits.

The article is organized as follows. Sec. **II** discusses the VOI. Sec. **III** explains the principle underlying the valley FET. Secs. **IV** and **V** discuss the valley pair qubit in the tilted magnetic field and in the in-plane field configurations, respectively. Last, Sec. **VI** summarizes the discussion.

**II. The VOI**

Graphene is an ideal system to illustrate the VOI. A free-standing graphene layer is gapless. For numerous applications, however, a great challenge posed is to open a gap in the material. The existence of a finite gap permits, in the applications envisioned here, the formation of graphene QDs or QWs by the electrostatic method commonly used in patterning semiconductor QDs or QWs.

Several important proposals have appeared to address the fabrication of gapped graphene, for example, graphene nanoribbons[18], epitaxial graphene on h-BN[24] or Ru,[25], graphene nanoroads[26], biased bilayer graphene[27], strained graphene[28], and so on. Throughout the article, we focus on the structure – gapped epitaxial graphene, and discuss the VOI in the structure. However, a recent theoretical study shows that the VOI exists also in biased bilayer systems.[29]

In epitaxial graphene, the graphene-substrate interaction breaks the sublattice symmetry, leading to the following Dirac type equation for electrons near K and K'

$$\begin{pmatrix} \Delta & v_F(i\hat{p}_x + \tau \hat{p}_y) \\ v_F(-i\hat{p}_x + \tau \hat{p}_y) & -\Delta \end{pmatrix} \begin{pmatrix} \psi_{A,\tau} \\ \psi_{B,\tau} \end{pmatrix} = (E + \Delta) \begin{pmatrix} \psi_{A,\tau} \\ \psi_{B,\tau} \end{pmatrix}. \qquad (1)$$

The energy E here is defined with respect to the conduction band edge. $(\psi_{A\tau}, \psi_{B\tau})^T$ is the "Dirac wave function" describing the probability amplitudes on sublattices A and B. $\Delta$ is the gap parameter determined by the graphene-substrate interaction. In the case where the substrate and graphene do not lattice match, a superstructure – Moire pattern is produced making Δ varying periodically in space.[30] We avoid this complication by assuming the application of a biaxial strain to match the lattices of graphene and substrate.

Eqn. (1) is basically the two-band model of gapped graphene in a field-free space, which treats both the conduction and valence bands simultaneously. From Eqn. (1), E is given by

$$(E+\Delta)^2 = \Delta^2 + v_F^2 \hbar^2 \vec{k}^2,$$

where **k** = electron wave vector. This represents the spectrum of a massive Dirac particle, with Δ being the "rest mass energy" and the Fermi velocity $v_F$ being the corresponding "light velocity" in graphene.

In the presence of an external potential V, one adds V to the diagonal elements of Eqn. (1). In this case, the set of coupled differential equations in Eqn. (1) generally constitute a difficult mathematical problem. However, for E << Δ, the analogy between a graphene electron and a Dirac massive particle suggests a more convenient, one-band approximation, which is the analogy of the Schrodinger equation given below[13]

$$\begin{aligned}
H\phi_\tau &\approx E_\tau \phi_\tau, \\
H &= H^{(0)} + H^{(1)}, \\
H^{(0)} &= \frac{\vec{p}^2}{2m^*} + V, \\
H^{(1)} &= -\frac{\vec{p}^4}{8m^{*2}\Delta} - \frac{\vec{p}^2 V}{8m^*\Delta} + H_{vo}, \\
H_{vo} &= \tau \frac{\hbar}{4m^*\Delta}(\nabla V \times \hat{p})_z,
\end{aligned} \quad (2)$$

where H is the "Schrodinger Hamiltonian", with $H^{(0)}$ being the "non-relativistic part" and $H^{(1)}$ the 1st-order "relativistic correction" (R.C.). $E_\tau$ is the energy for a τ-valley electron. $\phi_\tau$ is the corresponding one-component "Schrodinger wave function" and derives from the component $\psi_{A,\tau}$ of Dirac wave function by the linear transformation, $\phi_\tau = (1 + p^2 / 8m^*\Delta) \psi_{A,\tau}$. Specifically, $\phi_\tau$ is interpreted as a probability amplitude, with $|\phi_\tau|^2 \approx |\psi_{A,\tau}|^2 + |\psi_{B,\tau}|^2$ being the probability distribution of an electron in the graphene lattice. The potential energy V consists of the confinement potential (of the QDs in the case of valley pair qubits, or of the QWs in the case of valley FETs) and the gate potential applied to the structure. Notice the appearance of $H_{vo}$ – the VOI – in the Hamiltonian, as a result of the reduction to the one-band model. Most interestingly, as the expression $H_{vo}$ shows, the VOI is $\tau$ (i.e., valley)-dependent and controllable through V, giving rise to the possibility of valley manipulation via electrical gates.

Eqn. (2) treats approximately the low-energy conduction electron state. A similar approximation can be applied to the valence hole case, due to the electron-hole symmetry in graphene. Moreover, Eqn. (2) can be extended to cover the effect of a normal magnetic field, or an AC electric field, as well as the higher order R.C.s. The readers are referred to References 13 and 14 for details.

The electron valley d.o.f. can be manipulated by coupling it to external fields, e.g., a magnetic field with a $B_{normal}$ component, or a DC or AC in-plane electric field. Summarized below are the results drawn from previous studies of Eqn. (2) (and its extensions), to be used as the basis for the discussions in Secs. **III**, **IV**, and **V**.

(A) The "Rashba effect"

In the presence of an in-plane electric field, $H_{vo}$ produces an effective magnetic field and induces valley splitting. For illustration, we consider the case of a graphene QW grown in the x-direction and subject to the confinement potential ½ $m^*w_0^2y^2 - Dy^4$, with D being the strength of the quartic term in the potential. (The quartic term places a cap on the rise of the parabolic term with increasing |y|, simulating the case of a finite confinement potential.) We further assume a gap variation in the y-direction, e.g., $\Delta(y) = \Delta + \Delta'(y)$, with $\Delta'(y)$ being the variation. In principle, the variation can be realized as follows. In the case of substrate-grown graphene, one can make a trough in the substrate, and place the graphene layer on the substrate. The region of graphene on top of the trough is free-standing and gapless, while a gap is produced in the region of graphene next to the trough due to the graphene-substrate interaction. Thus, a piecewise constant $\Delta(y)$ (or Δ'(y)) is generated. In the presence of Δ'(y), it can be shown that the VOI given in Eqn. (2) is generalized as follows, e.g.,

$$H_{vo} \to H_{vo}' = H_{vo} - \tau \frac{\hbar}{4m^{*}\Delta} \left[\nabla \Delta'(y) \times \hat{p}\right]_z.$$

Fig. 1 shows the splitting of the lowest subbands induced by a lateral electric field, due to the VOI. For a given electron energy E, the splitting leads to the following wave vector difference[16]

$$k_+ - k_- = \frac{2m^{*}\alpha_{vo}}{\hbar^2}, \qquad (3)$$

$$\alpha_{vo} = \frac{6e\hbar^3}{m^{*4}\Delta_0 w_0^5} D \delta \, \varepsilon_y,$$

$k_+$ ($k_-$) being the wave vector of the electron with $\tau = +$ (-). Here, $\varepsilon_y$ = the lateral electric field in the y-direction, and Δ'(y) is modeled with a quadratic function, e.g., Δ'(y) = δ $y^2$, where δ << $m^{*}w_0^2$ (i.e., weak gap variation). The splitting shown in Eqn. (3) is similar to the Rashba effect in semiconductors, and the magnitude of splitting here is determined by $\alpha_{vo}$, the "Rashba constant". A reasonable estimate gives $\alpha_{vo} \approx 5.4 \times 10^{-12} eV \cdot m$ [16], comparable to the Rashba constant in semiconductors with large SOI, e.g., InAs.[31]

**(B) Electrical tuning of valley magnetic moment**

In the case of a graphene QD, an in-plane electric field can be applied to modulate the valley magnetic moment $\mu_v$ of a confined electron, as follows. Let the QD confinement potential be symmetric and given by ½ m* $w_0^2$ ($x^2 + y^2$) – ¼ $k_4$ m* $w_0^2$ ($x^4 + y^4$), with the quartic term again limiting the increase of the quadratic term. For an electron in the ground state, $\mu_v$ is given by[13]

$$\mu_v = \mu_{v0}\left[1 - \frac{\hbar w_0}{4\Delta}\right], \qquad (4)$$

$$\mu_{v0} \equiv \frac{e\hbar}{2m^{*}}.$$

Here, $\mu_{v0}$ is the nonrelativistic part of the moment, and is an analogue of the Bohr spin magneton. Notice the modification of $\mu_{v0}$ by the additional term in Eqn. (4), which derives from $H^{(1)}$ and is proportional to the ratio "$\hbar w_0 / \Delta$" (i.e., electron energy to rest mass energy, with "$\hbar w_0$" being the electron energy), as is typical of a R.C. This R.C. gives the access to the tuning of $\mu_v$ by electrical means – by shifting the electron energy with an electrical field (the Stark effect) to produce a change in "$\hbar w_0 / \Delta$". The corresponding variation $\delta\mu_v$ generated in the magnetic moment is given by[13]

$$\delta\mu_v \approx \mu_{v0}(\frac{3}{16}k_4 y_\varepsilon^2)\frac{\hbar w_0}{\Delta}, \tag{5}$$
$$y_\varepsilon \equiv e\varepsilon_y / m^* w_0^2,$$

where $\varepsilon_y$ = the electric field applied.

### (C) AC electric field-induced valley asymmetry

Just as a magnetic field with the $B_{normal}$ component can break the valley symmetry / degeneracy, so can an AC electric field. We consider the same symmetric QD as specified in **(B)**. For the discussion here, we take the AC electric field to be $\varepsilon_y \sin(w_s t)$ applied in the y-direction. Furthermore, the QD is made reflection asymmetric with respect to the y-axis, by the application of a DC electric field ($\varepsilon_x$) in the x-direction. Ref. 14 shows that the AC field breaks the valley symmetry and induces valley-contrasting phase evolutions, which are given by

$$|K> \to |K> e^{i\Phi_{1/2}}, \tag{6}$$
$$|K'> \to |K'> e^{-i\Phi_{1/2}},$$
$$\Phi_{1/2} = y_\varepsilon / l_{v0}.$$

Here,

$$y_\varepsilon = e\varepsilon_y / m^* w_0^2, \quad l_{vo}^{-1} = \frac{1}{12}\frac{\hbar^2 w_0^2}{\Delta^2}k_4 x_\varepsilon, \quad x_\varepsilon = e\varepsilon_x / m^* w_0^2.$$

$l_{vo}$ is called the valley-orbit length.

### III. Valley FETs

Fig. 2 shows the structure of a valley FET with a side gate, AGNRs as the leads, and a graphene QW as the channel. Let $L$ = channel length, and $W$ = AGNR width, with the nanoribbon edges located at y = ±$W/2$. We discuss electron states in the lead and in the channel, and the VOI-based conductance modulation for the on-off switch function of the FET.

lead states

For an electron in the lead, the K and K' valleys are mixed. The total Dirac wave function is subject to the boundary condition $\psi_A(y = W/2) = \psi_B(y = W/2) = \psi_A(y = -W/2) = \psi_B(y = -W/2) = 0$. Let $k$ = electron wave vector along the AGNR. The corresponding wave function satisfying the boundary condition is given by

$$\begin{pmatrix}\Psi_A \\ \Psi_B\end{pmatrix} \propto \left[\begin{pmatrix}e^{ikx}e^{i\vec{K}\cdot\vec{r}} & S_{K'/K}e^{ikx}e^{i\vec{K}'\cdot\vec{r}}\end{pmatrix}\begin{pmatrix}e^{ik_y y} \\ e^{-ik_y y}\end{pmatrix}\right]\begin{pmatrix}1 \\ \hbar v_F(k_y - ik) \\ \overline{2\Delta + E}\end{pmatrix}. \qquad (7)$$

"**K**" and "**K'**" in the equation are the wave vectors at K and K' points, respectively. $k_y$ (lateral wave vector) and E (electron energy) are both quantized due to the lateral quantum confinement in AGNR. $S_{K'/K}$ is the amplitude of K' component relative to that of K component, with $|S_{K'/K}| = 1$. Therefore, the source / drain state in Eqn. (7) is valley-mixed in the 50-50 ratio. This specific "valley polarized state" is utilized for valley injection / detection.

channel states

The QW is subject to the same confinement potential and gate field as specified in **(A)**, Sec. **II**. The electron state is solved approximately within the Schrodinger model (i.e., Eqn. (2)). It gives the channel state of an injected electron as a linear combination of K and K' components, with the following real space wave function

$$\Phi_0 \approx \begin{pmatrix}e^{ik_+ x}e^{i\vec{K}\cdot\vec{r}} & C_{K'/K}e^{ik_- x}e^{i\vec{K}'\cdot\vec{r}}\end{pmatrix}\begin{pmatrix}\exp(-\beta y^2) \\ \exp(-\beta y^2)\end{pmatrix}. \qquad (8)$$

$\beta = m^* w_0 / 2\hbar$. The Gaussian function here is the simple harmonic oscillator ground state. The parameter $C_{K'/K}$ in $\Phi_0$ is chosen to match the valley polarization of channel state with that of the source state (as specified in Eqn. (7)) at the source / channel interface (located at x = 0), giving $C_{K'/K} = S_{K'/K}$.

<u>conductance modulation</u>

As discussed above, the initial channel state (at x = 0) matches the source state in valley polarization. However, because the K and K' components of $\Phi_0$ travel with separate wave vectors ($k_+$ and $k_-$), due to the Rashba effect given in Eqn. (3), they evolve with difference phases. Therefore, the electron starts precession in the valley space as the electron moves away from x = 0. Specifically, when it arrives at the channel / drain interface (located at x = L), the phase difference accumulated over the trip is given by

$$\delta\varphi = \frac{2m^* \alpha_{vo}}{\hbar^2} L \qquad (9)$$

between the two valley components. Combining Eqns. (3) and (9), it gives $\delta\varphi \propto \varepsilon_y$, meaning that $\delta\varphi$ is controllable by applying a gate bias.

$\delta\varphi$ determines the valley polarization of the channel state at x = L relative to that of the drain state. For $\delta\varphi = 2m\pi$ ($\delta\varphi = (2m+1)\pi$), the channel and drain polarizations are aligned with (orthogonal to) each other, leading to a conductance maximum (minimum). This achieves the on-off switch function required of a FET.

The valley FET is an anologue of the Datta-Das spin FET[32]. By its analogy to the spin FET, the valley FET provides a potential framework to build low-power FET circuits for graphene-based nanoelectronics. The following table shows the correspondence between the two FETs

| **FET** | **d.o.f.** | **lead** | **Q1D channel** | **electrical control** | **physical mechanism** |
|---|---|---|---|---|---|
| valley FET | valley K,K' | AGNR | graphene | side gate | valley-orbit interaction (VOI) |
| spin FET | spin ↑,↓ | FM | semiconductor | top gate | Rashba SOI |

However, there are qualitative differences. A valley FET is characterized by: i) smooth lead / channel interfaces, in contrast to the FM / semiconductor interface (in a spin FET) which presents the challenge of conductivity mismatch[33] and requires a structure modification such as the insertion of tunnel junctions[34], and ii) vanishing interband flip type scattering, in contrast to the interband spin flip scattering (in a spin FET) which gives rise to random conductance oscillations[35] and requires, for example, the introduction of stray electric fields into the device[36]. Therefore, the valley FET provides an ideal implementation of the original Datta-Das idea in the graphene-based electronics.

## IV. Valley pair qubits in the tilted magnetic field configuration

Fig. 3 shows the structure of a valley pair qubit, which consists of a pair of laterally coupled QDs placed in a tilted magnetic field. The Fermi level is set for two electrons confined in the structure. Several gates are utilized for the qubit manipulation. The central one controls the barrier between the QDs, and the side gates control the QD energy levels and valley magnetic moments, as described in **(B)**, Sec. **II**.

In essence, the state of a valley pair qubit (0 or 1) is a low-energy, two-electron quantum state, where the electrons are separately confined in the QDs. In the theoretical model of the qubit, the spin d.o.f. is frozen by the magnetic field, as shown in Fig. 4. The orbital d.o.f. is frozen, too, into the QD ground state, except for the tunneling characterized by $t_{d-d}$ (tunneling energy) between the QDs. In addition, we allow for the on-site Coulomb interaction, with U = on-site Coulomb energy. The low energy states of the two-electron system are "valley singlet" $|Z_S\rangle$ and "valley triplets" $\{|Z_{T0}\rangle, |Z_{T+}\rangle, |Z_{T-}\rangle\}$, given by

$|Z_S\rangle = (½)^{1/2} (c_{KL}^+ c_{KR'}^+ - c_{KL'}^+ c_{KR}^+) |\text{vacuum}\rangle$,

$|Z_{T0}\rangle = (½)^{1/2} (c_{KL}^+ c_{KR'}^+ + c_{KL'}^+ c_{KR}^+) |\text{vacuum}\rangle$,

$|Z_{T+}\rangle = c_{KL}^+ c_{KR}^+ |\text{vacuum}\rangle$,

$|Z_{T-}\rangle = c_{KL'}^+ c_{KR'}^+ |\text{vacuum}\rangle$.

$c / c^+$ are the electron creation / destruction operators, and the subscripts L / R denote the ground states in the left / right QDs. Valley singlet and triplet states are split by J, with $J \sim 4 t_{d-d}^2/(U-\delta\varepsilon)$ being the exchange coupling. $\delta\varepsilon$ here refers to the energy detuning between the QD levels. Note that because J is $t_{d-d}$ dependent, it is controllable by the central gate in the structure.

In the scheme of valley pair qubits, $|Z_S\rangle / |Z_{T0}\rangle$ represent logical 0 / 1. $|Z_{T+}\rangle$ and $|Z_{T-}\rangle$ are redundant and not used in quantum information processing. However, $|Z_{T+}\rangle$ and $|Z_{T-}\rangle$ provide

leakage channels in the presence of impurity or confinement potential -induced intervalley scattering. The associated decoherence rate has been estimated and found to be low enough for qubit manipulation.[13,14] Generally, the valley pair scheme resembles the spin pair scheme[37-40] which uses the spin singlet / triplet states as 0 / 1. Therefore, it shares the unique advantage of the spin pair scheme – being decoherence-free in the space of qubit states (expanded by $|Z_S\rangle$ and $|Z_{T0}\rangle$ and denoted as $\Gamma$ in the valley pair scheme). Moreover, because of the resemblance, the method developed for initialization / readout / qugate operation in the spin pair scheme[40] can be adapted here. The following discussion focuses on the manipulation of a single qubit.

In the space $\Gamma$, the qubit model is chracterized by the following effective Hamiltonian

$$H_{eff} = (\mu_{vL} - \mu_{vR})B_{normal}\tau_x + \frac{J}{2}\tau_z \qquad (10)$$

$\mu_{vL}$ and $\mu_{vR}$ are the electron valley moments in the left and right QDs, respectively. $\tau_x$ and $\tau_z$ are the standard Pauli matrices. The corresponding eigenstates of $H_{eff}$ are shown in Fig. 5, where $|X_+\rangle \equiv c_{KL}^+ c_{KR'}^+|vacuum\rangle$, $|X_-\rangle \equiv c_{KL'}^+ c_{KR}^+|vacuum\rangle$.

$\Gamma$ is isomorphic to the spin-½ state space, and Eqn. (10) is analogous to the interaction between a spin-1/2 electron and a magnetic field, e.g., "$\mathbf{s} \cdot \mathbf{B}$", with $\mathbf{B} = (B_x, B_z)$. The analogy is summarized in the following table:

|  | qubit / spin states | magnetic field components |  |
|---|---|---|---|
| qubit | $|Z_S\rangle, |Z_{T0}\rangle, |X_+\rangle, |X_-\rangle$ | $(\mu_{vL} - \mu_{vR})B_{normal}$,   $J/2$ | (T-1) |
| spin-1/2 | ↑, ↓, →, ← | $B_x$,   $B_z$ |  |

Here, ↑, ↓, →, and ← are the spins aligned in the +z, –z, +x, and –x directions, respectively. By the analogy in (T-1), it is immediately clear about how to manipulate the qubit into any state in $\Gamma$. To envision this, we employ the Bloch sphere description of qubit / spin states, where each qubit / spin state is represented by a point on the sphere. "$(\mu_{vL} - \mu_{vR})B_{normal}$" / "$B_x$" produces $R_x$, a rotation of qubit / spin about the x-axis of the sphere. "$J/2$" / "$B_z$" produces $R_z$, a rotation about the z-axis. Combined together, these two independent rotations are sufficient to move the qubit / spin from one point to any other point on the sphere, as shown in Fig. 6. Moreover, since J and $\mu_{vL(R)}$ are all controllable via electrical gates (with the former by the central gate, and the latter by the side gates as prescribed in **(B)**, Sec. **II**), it provides an arbitrary, all-electrical manipulation of a single qubit. The single-qubit manipulation and a qugate operation together constitute a universal quantum computation.[41-43]

It is interesting to compare the implementations of the valley pair qubit and the spin pair qubit. While they are both based on similar principles, e.g., Eqn. (10), there is a significant difference between the two, in the method employed to generate an $R_x$. In the valley pair scheme, "($\mu_{vL} - \mu_{vR}$)" is controllable by the side gate voltages, while the spin pair scheme requires a challenging, materials-based "g-factor engineering" for the creation of the gradient. The "gate-based g-factor engineering" for $R_x$ in the valley pair scheme is an attractive feature from the implementation point of view.

We make a remark here on the role of $B_{normal}$. The presence of $B_{normal}$ breaks the time-reversal symmetry and lifts the valley degeneracy in graphene. In addition to its utilization in generating an effective $B_x$ (for qubit manipulation), the valley splitting produced between |K> and |K'> forbids the valley flip scattering, |K> ↔ |K'>, unless phonons are involved to compensate for the energy difference. This suppresses, at low temperatures, the flip scattering and hence the leakage from $\Gamma$ into the redundant triplet states $Z_{T+}$ and $Z_{T-}$.

### V. Valley pair qubits in the in-plane magnetic field configuration

Valley pairs qubits can also work in an in-plane magnetic field, which is the configuration required for quantum communications. We shall explain this point before we move to the discussion of qubit manipulation in this configuration.

In long-distance photonic quantum communications, apart from the flying photon qubits which carry quantum information between locations, static qubits serve the important function as quantum memories, in the quantum repeater protocol required in the communications.[44-46] There are various forms of memory qubits, among which graphene-based ones, such as valley pair qubits considered here, are good choices, because of the potential to make them into gated structures for electrical control.

In any form of memory qubits, it is important to require good fidelity in the quantum state transfer (QST) between photon and memory qubits, an elementary process which occurs frequently in the quantum repeater-based photonic quantum communications. For the application of valley pair qubits as quantum memories, we consider specifically the case where the quantum information in a photon is encoded in the state of circular polarization, e.g., $\alpha|\sigma+\rangle + \beta|\sigma-\rangle$. A QST from the photon qubit to a valley qubit occurs via the absorption of the photon by the graphene QDs in the valley qubit, and obviously, a faithful QST, e.g., $\alpha|\sigma+\rangle + \beta|\sigma-\rangle \rightarrow \alpha|K_L K'_R\rangle + \beta|K'_L K_R\rangle$, requires a valley-symmetric optical response of the QDs with respect to the polarization.[14] Explicitly, it means, in a QD, i) that $|K\rangle$ and $|K'\rangle$ must be degenerate; and ii) that the matrix elements for the optical transitions $|\sigma+\rangle + |K(\text{valence})\rangle \rightarrow |K(\text{conduction})\rangle$ and $|\sigma-\rangle + |K'(\text{valence})\rangle \rightarrow |K'(\text{conduction})\rangle$ must be the same in magnitude. (A phase difference in the complex-valued matrix elements does not cost fidelity. It generates an extra phase in $\alpha$ relative to $\beta$, which can be removed by a pre-QST or post-QST qubit manipulation.) Both of the conditions are tied to the time-reversal symmetry and, hence, are fulfilled only if $B_{normal} = 0$. On the other hand, an in-plane $B_{plane}$ is still needed to freeze the spin d.o.f. of electrons.

It should be noted that, in the communications envisioned here, apart from the frequent QST – photon qubit $\rightarrow$ valley pair qubit, the reverse process – valley pair qubit $\rightarrow$ photon qubit occurs just as frequently. In fact, the two processes have to alternate with each other in long distance communications, because the quantum information is stored only in one form of the qubits – either photon or valley pair qubits, in each stage of the communication. Interestingly, theoretical analysis[14] finds that while there is some degree of residual, intrinsic distortion of the quantum information in each QST (meaning that in the QST, the coefficients $\alpha$ and $\beta$ in the quantum state of one qubit do not fully carry over to the state of the other qubit, even with the access to pre-QST or post-QST qubit manipulations available to treat the distortion), the composite back-to-back QST, namely, valley $\rightarrow$ photon $\rightarrow$ valley, can reach 100% intrinsic fidelity, due to the exact cancellation of quantum distortions between the two individual QSTs. The same conclusion applies to similar but longer processes, e.g., valley $\rightarrow$ photon $\rightarrow$ valley $\rightarrow$ …… $\rightarrow$ valley, barring the external factors such as those leading to photon attenuation or phonon-involved qubit decoherence.

The above discussion shows the potential of graphene quantum dots in photonic quantum communications. Along with the previous discussion (in Sec. **IV**) of valley pair qubits for quantum computing, it suggests the interesting prospect of an all-graphene approach for the solid state components of a quantum network, i.e., quantum computers and quantum memories in communications, as shown in Fig. 7.

Now, we discuss the qubit manipulation. In the in-plane magnetic field configuration, $B_{normal}$ is not available for the generation of $R_x$. However, as explained below, AC electric fields can be applied through the side gates ($V_L$ and / or $V_R$) to produce the same rotation. Based on the result for a single-QD state in **(C)**, Sec. **II**, the qubit states evolve as follows,

$$\begin{array}{ll} |K_L K_R'\rangle \to e^{i\theta_x/2} |K_L K_R'\rangle \\ |K_L' K_R\rangle \to e^{-i\theta_x/2} |K_L' K_R\rangle \end{array}, \text{ or } \begin{array}{l} |X_+\rangle \to e^{i\theta_x/2} |X_+\rangle \\ |X_-\rangle \to e^{-i\theta_x/2} |X_-\rangle \end{array}, \quad (11)$$

$$\frac{\theta_x}{2} = \frac{y_{\varepsilon,L}}{l_{vo,L}} - \frac{y_{\varepsilon,R}}{l_{vo,R}},$$

in half of the AC cycle. $y_{\varepsilon,L(R)}$ and $l_{vo,L(R)}$ are $y_\varepsilon$ and $l_{vo}$ introduced in Eqn. (6) for a QD, defined here with respect to the left and right QDs, respectively. The contrasting phase evolutions in $|X_+\rangle$ and $|X_-\rangle$ shown here represents a rotation $R_x$. On the other hand, a rotation $R_z$ can be generated by the exchange coupling, in exactly the same fashion as discussed previously in Sec. **IV**. These two rotations are combined to furnish any necessary single-qubit manipulation, all-electrically.

## VI. Summary

We have discussed the concept of valley-orbit interaction based valleytronics, which can be applied to, but not limited to, graphene-based electronics. Two important examples based on the VOI mechanism have been illustrated, namely, 1) valley pair qubits to build quantum networks consisting primarily of graphene and photons, and 2) valley FETs to build graphene-based, low-power FET circuits.

In the case of gapped graphene, the unique physical mechanism of VOI is characterized by (i) a large strength; (ii) a full potential for electrical control of the valley d.o.f. with standard electrical gate operations; and (iii) valley-conservingness. These facts make the VOI-based valleytronics an attractive R & D direction in the area of electronics.

**Acknowledgment** – We would like to thank the support of ROC National Science Council through the contract No. NSC100-2112-M-007-009.

**Figure captions**

**Fig. 1** The lowest subbands $E_{0,\tau}(k_\tau)$ for $\tau = \pm$ in the graphene QW. For a given energy E, the subbands for K and K' are horizontally split due to the Rashba effect.

**Fig. 2** The VFET shown as a three-terminal device with a side gate. The source and drain are AGNR's, and the channel is a graphene quantum wire (G-QW) aligned in the armchair direction. The source and drain inject and detect electrons in a specific polarization. When an electron moves down the channel, the valley precession occurs due to the gate field-induced VOI.

**Fig. 3** The structure of a valley pair qubit. The QDs are electrostatically defined, for example, by back gates (not shown in the figure). Gate $V_C$ controls the potential barrier between the QDs (and hence the tunneling coupling $t_{d-d}$ between the two QDs), and also provides the electric field $\varepsilon_x$ in each QD, mentioned in **(C)**, Sec. **II**. Gates $V_L$ and $V_R$ control the valley magnetic moments in the QDs, or may be applied with AC voltages.

**Fig. 4** One-electron energy levels in a graphene QD, where a static, tilted magnetic field is applied. For the two-electron valley singlet / triplet states, however, only the lowest levels of K and K' valleys in each QD are occupied. The spin d.o.f. ($\sigma$) is frozen at $\sigma = +½$ due to the magnetic field.

**Fig. 5** Valley singlet ($z_S$) / triplet ($z_{T0}$, $z_{T+}$, $z_{T-}$) states constitute the low energy sector of the two-electron states. The singlet-triplet splitting is given by "J", for $\mu_{vL} = \mu_{vR}$ (left), while that between $|X_+\rangle$ and $|X_-\rangle$ by "$2(\mu_{vL} - \mu_{vR})B_{normal}$", for J = 0 (right).

**Fig. 6** The time evolution of a qubit state in the Bloch sphere, as governed by $H_{eff}$. $\check{R}_x(\theta_x)$ represents a rotation about the x-axis, and $\check{R}_z(\theta_z)$ about the z-axis, of the Bloch sphere. $\theta_x$ and $\theta_z$ are the respective angles of rotation, determined by the effective magnetic fields "$(\mu_{vL} - \mu_{vR})B_{normal}$" and "J/2", respectively.

**Fig. 7** A schematic plot of the quantum network consisting of graphene and photons.

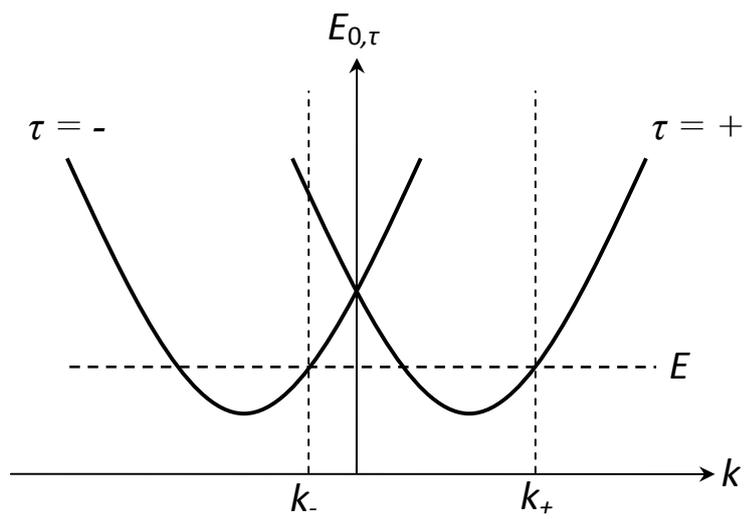

**Fig 1**

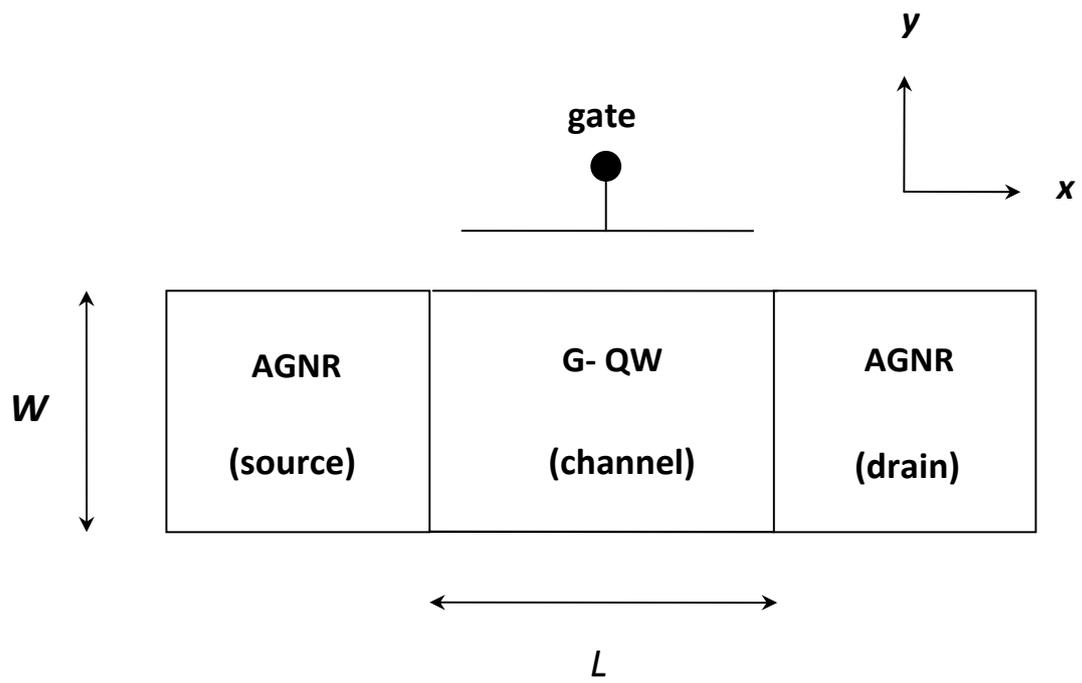

**Fig 2**

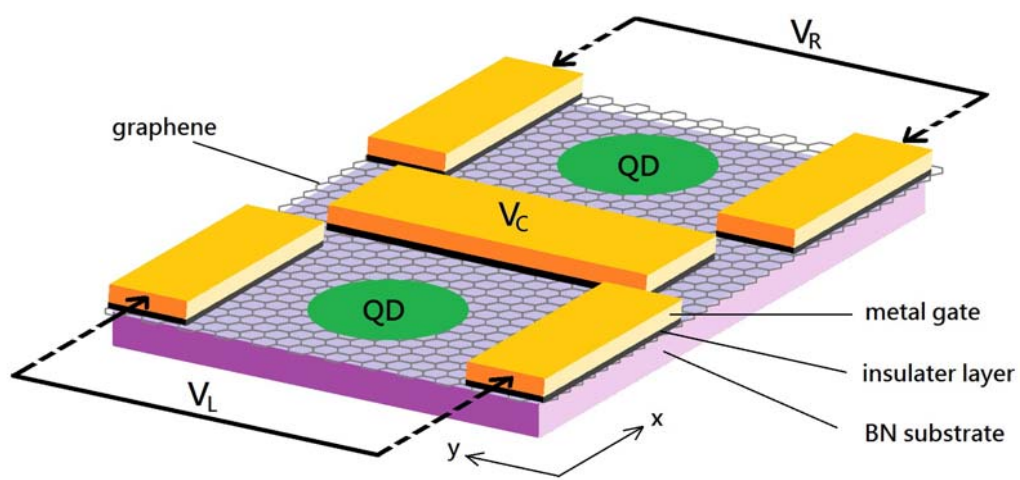

**Fig 3**

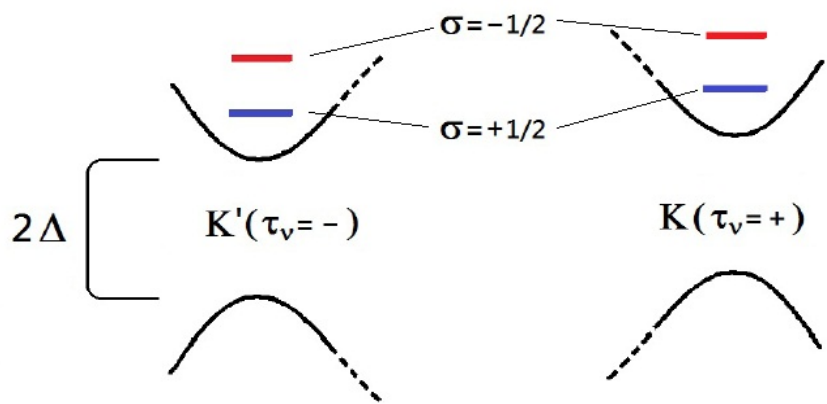

**Fig 4**

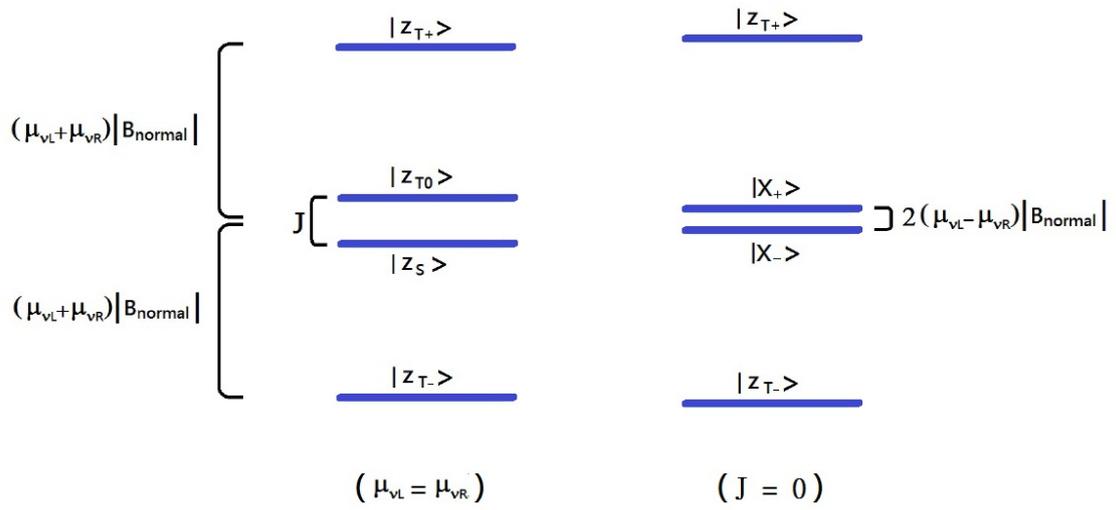

**Fig 5**

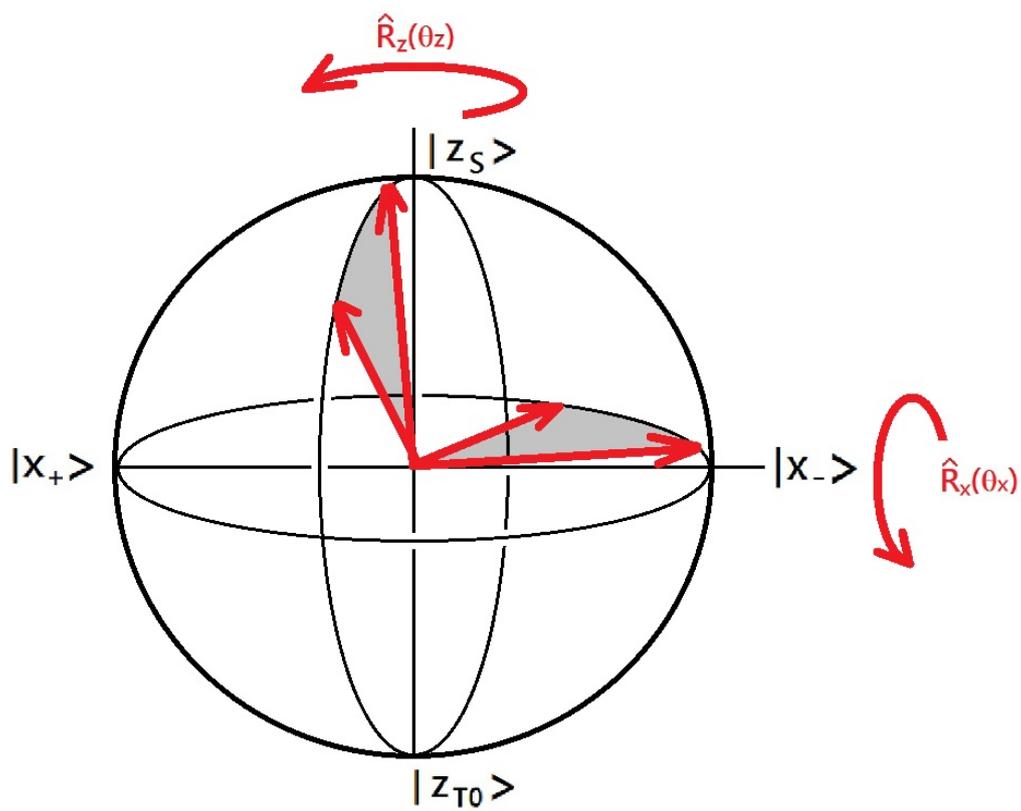

**Fig 6**

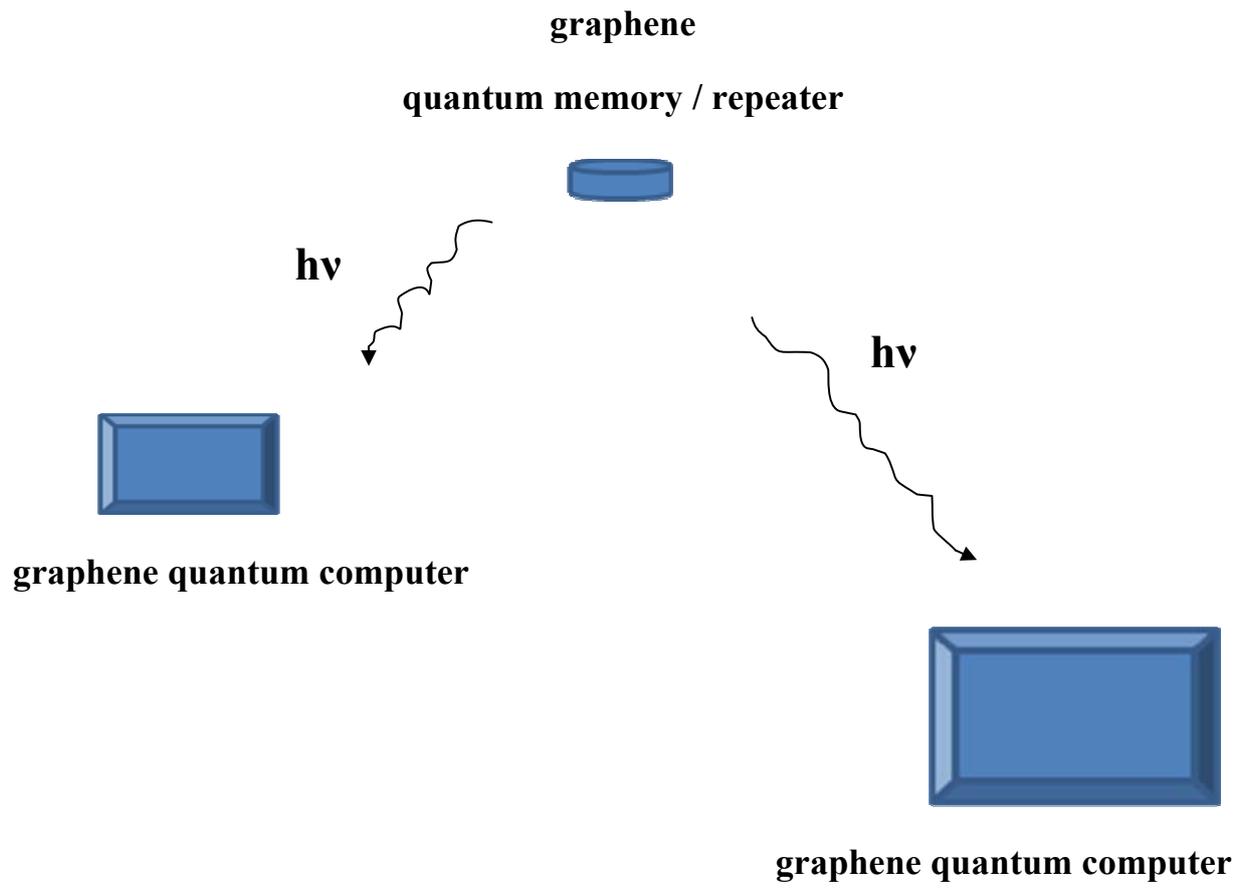

**Fig 7**